# Measurements of the Influence of Acceleration and Temperature of Bodies on their Weight

## A. L. Dmitriev

*St-Petersburg State University of Information Technologies, Mechanics and Optics*
*49, Kronverksky Prospect, St-Petersburg, 197101, Russia, dalexl@rol.ru*

**Abstract.** A brief review of experimental research of the influence of acceleration and temperatures of test mass upon gravitation force, executed between the 1990s and the beginning of 2000 is provided. According to a phenomenological notion, the acceleration of a test mass caused by external action, for example electromagnetic forces, results in changes of the gravitational properties of this mass. Consequences are a dependence upon gravity on the size and sign of test mass acceleration, and also on its absolute temperature. Results of weighing a rotor of a mechanical gyroscope with a horizontal axis, an anisotropic crystal with the big difference of the speed of longitudinal acoustic waves, measurements of temperature dependence of weight of metal bars of non-magnetic materials, and also measurement of restitution coefficients at quasi-elastic impact of a steel ball about a massive plate are given. A negative temperature dependence of the weight of a brass core with relative size near $5 \cdot 10^{-4}$ K$^{-1}$ at room temperature was measured; this temperature factor was found to be a maximum for light and elastic metals. All observably experimental effects, have probably a general physical reason connected with the weight change dependent upon acceleration of a body or at thermal movement of its microparticles. Paper presented at the 5-th Symposium on New Frontiers and Future Concepts (STAIF-2008), Albuquerque, New Mexico 10 – 14 Feb. 2008.

## INTRODUCTION

The deep interrelation of electromagnetic and gravitational forces can and should be manifested in experiments with exact weighing of the test bodies moving with acceleration under action of elastic forces. For a long time the attention given to this problem was unduly not enough which was in part promoted by theoretical concepts of the general relativity regarding fictitiousness of the «force of gravitation» concept. Meanwhile, a number of experimental measurements have tested the influence of acceleration of external elastic forces on the value of acceleration of gravity as it will be shown in the present paper.

We propose to characterize this influence as follows. If a test body under action of external elastic forces moves upwards with acceleration value $a_g$, an increment $\Delta g_c$ of the gravitational acceleration shall occur, which is given in a first linear approximation as

$$\Delta g_c = \alpha_c a_g \quad . \tag{1}$$

Similarily, if a body is accelerated downwards under the action of external forces, then an increment $\Delta g_p$ of the gravitational acceleration shall occur with a different sign, given by

$$\Delta g_p = -\alpha_p a_g \ . \tag{2}$$

In Equ. (1) and Equ. (2) the dimensionless factors $\alpha_c$ and $\alpha_p$ characterize a degree of interaction of elastic and gravitational forces. Vertical harmonious oscillations of a test body with a weight p will therefore lead to an average weight given by

$$p = mg_0 \left[ 1 - \frac{(\alpha_p - \alpha_c) A \omega^2}{\pi g_0} \right] \ , \tag{3}$$

where m is the mass of a body, $g_0$ the standard acceleration of gravity, A the amplitude, and $\omega$ the circular frequency of oscillations (Dmitriev, 2001).

The experiment should give an answer to a question whether the factors $\alpha_c$ and $\alpha_p$ are different from zero, and what their ratio is.

## MEASUREMENTS OF INFLUENCE OF ACCELERATION ON GRAVITY

The simple way of estimation of difference value ($\alpha_p$ - $\alpha_c$) is based on weighing of a rotor of a mechanical gyroscope with a horizontal axis of rotation. The role of elastic forces is played here by centripetal forces. It is possible to show that the weight P of the rotor in form of a cylinder with radiuses $R_1$ and $R_2$ is equal to

$$P = Mg_0 \left[ 1 - (\alpha_p - \alpha_c) \frac{2(R_2^3 - R_1^3)}{3\pi g_0 (R_2^2 - R_1^2)} \omega^2 \right] \ , \tag{4}$$

where $\omega$ is the angular speed of rotation.

Such an experiment was executed in 1999-2000 at Saint Petersburg (Dmitriev and Snegov, 2001). In this case, we weighed a pair of coaxial rotors rotating in opposite directions for compensation of the total angular moment of the container ($R_1$=15 mm, $R_2$=25 mm, M=250 g), as shown in Fig. 1. The obtained experimental dependence is shown in Fig. 2.

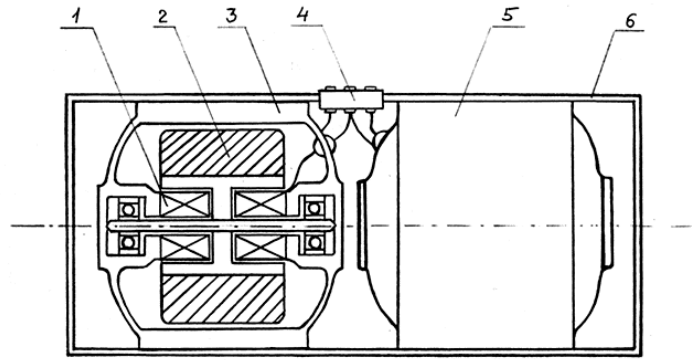

**FIGURE 1.** The device of the container. 1 - electric coils of the engine of a gyroscope, 2 - a massive cylindrical part of a rotor, 3 - the case of the first gyroscope, 4 - plugs of power supplies of engines of gyroscopes, 5 - the case of the second gyroscope (it is shown without a section), 6 - the case of the container

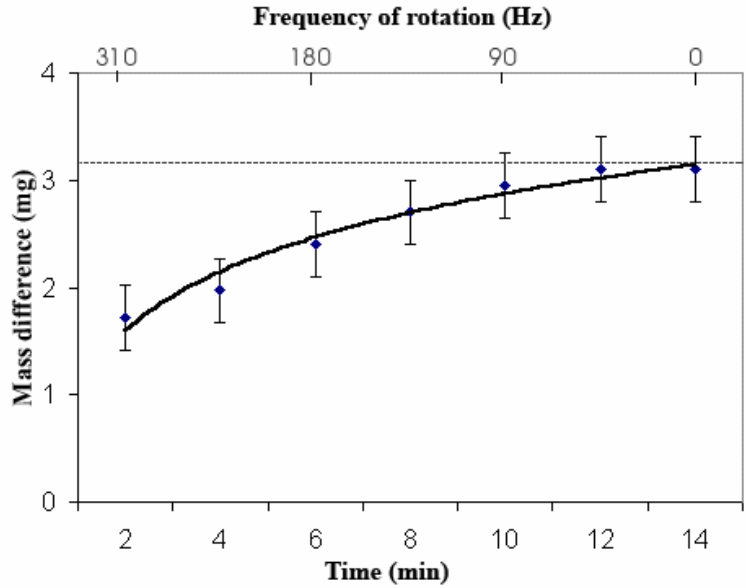

**FIGURE 2.** Mass difference of a horizontal and vertical rotor.

At a speed of rotation of 18.6 thousand rev/min the relative reduction of weight of a rotor was equal to $3·10^{-6}$. The estimated value of $(\alpha_p - \alpha_c)$ is near to $10^{-7}$. The factor $\alpha_c$ alone was evaluated by precision measurements for the restitution coefficients of an elastic impact of a ball against a massive metal plate. In these experiments a plate (and a ball trajectory) took horizontal and vertical positions (Dmitriev, 2002). Acceleration of the ball during impact duration exceeded of $10^4·g_0$. The difference of restitution coefficients in vertical ($k_1$) and horizontal ($k_2$) quasi-elastic impacts of the ball is shown in Fig. 3.

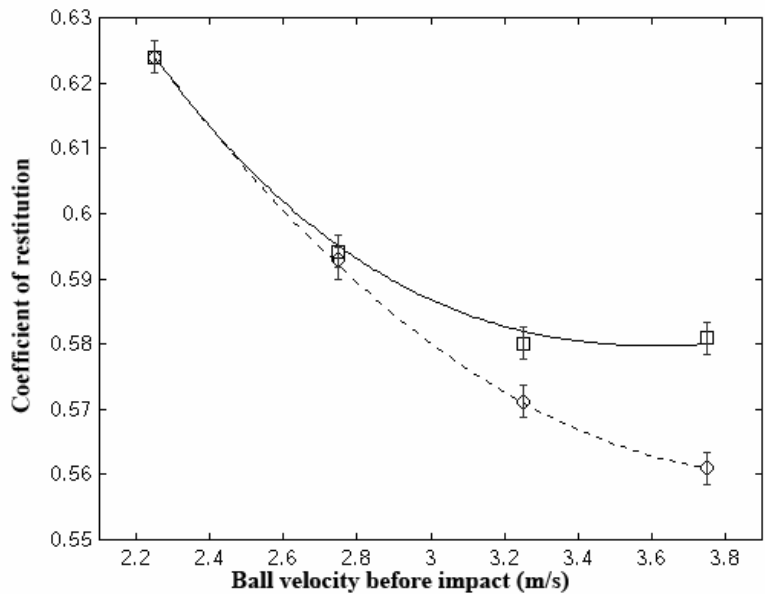

**FIGURE 3.** Experimental dependence of coefficient of restitution; the top line – $k_2$, the bottom line – $k_1$.

The order of value of the factor $\alpha_c$ can be estimated by the formula

$$\alpha_c \approx \frac{k_2 - k_1}{1 + k_2} \quad . \tag{5}$$

The speed of the ball before impact is about 3.5 m/s, which gives a $|\alpha_c| \cong 10^{-2}$ that is unexpectedly big.

Interesting results obtained by M. Tajmar's group in experiments with a rotating superconductor (Tajmar el al, 2007) probably have a physical nature close to the one discussed in the present work.

## MEASUREMENTS OF TEMPERATURE DEPENDENCE OF BODY WEIGHT

If indeed there is an influence of acceleration of elastic (electromagnetic in nature) forces on gravitation, then there will be the temperature dependence of body weights due to the thermal movements inside the body. The acceleration of microparticles in their thermal movement directly depend on their energy, and therefore from the absolute temperature of body. It is possible to show that in a classical approximation at temperatures higher than the Debye-temperature, the temperature dependence of body weight is described by the formula

$$P = Mg_0 \left[ 1 - \frac{(\alpha_p - \alpha_c)}{\pi g_0} C\sqrt{T} \right], \tag{6}$$

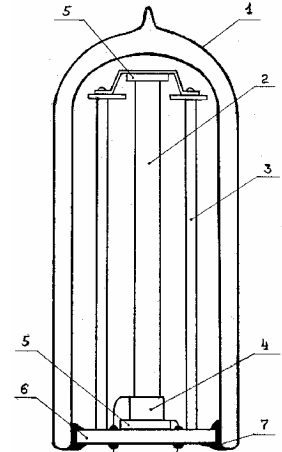

where C is a the factor dependent on physical characteristics (including density and elasticity) of bodies and T is the absolute temperature.

**FIGURE 4.** 1. Layout of hermetically sealed container. 1 – Dewar flask; 2 – metal rod; 3 – holder support; 4 – electroacoustic transducer (PZT); 5 – gaskets (foam plastic); 6 – holder base (ebonite); 7 – cold welding.

According to Equ. (6), an increase in the absolute body's temperature will cause a reduction of its weight. Such an effect was indeed observed in exact weighing of metal samples from nonmagnetic materials heated with ultrasound (Dmitriev, Nikushchenko and Snegov, 2003).

The layout of the hermetically sealed container shown in Fig. 4. An example of the experimental dependence of a sample weight in the process of its heating and cooling is shown in Fig. 5.

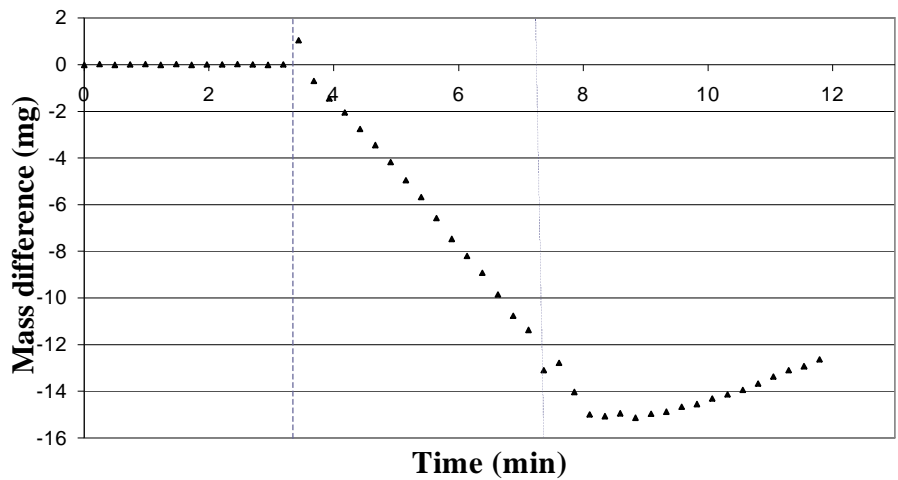

**FIGURE 5.** Change in mass of a brass rod. Ultrasound frequency 131.25 kHz. The touch lines indicate the moments when the ultrasound was switched on and off.

The temperature dependence of weight of various samples made of lead, copper, brass, titan and duralumin was measured. Some results of measurements are shown in Table 1.

**TABLE 1.** Characteristics of Samples and Results of Measurement, where $a=(\alpha_p-\alpha_c) \cdot C / \pi g_0$ and temperature factor $\alpha=-a/2\sqrt{T}$.

| Sample | Lead | Copper* | Brass** | Titanium | Duralumin |
|---|---|---|---|---|---|
| Length (mm) | 80.2 | 71.6 | 140.0 | 140.0 | 140.0 |
| Diameter (mm) | 8.0 | 10.5 | 8.0 | 8.0 | 8.0 |
| Mass (g) | 45.6 | 39.2 | 58.5 | 31.2 | 19.1 |
| Ultrasound Frequency (kHz) | 135.43 | 129.70 | 131.27 | 136.22 | 134.90 |
| $\Delta m/\Delta t$ (mg/min) | 1.06 | 1.15 | 2.64 | 1.63 | 1.66 |
| $\Delta T/\Delta t$ (K/min) | 5.1 | 4.5 | 10.0 | 6.0 | 7.5 |
| $\alpha \cdot 10^6$ (K$^{-1}$) at $T \cong 300$ K | 4.56 | 6.50 | 4.50 | 8.70 | 11.60 |
| $a \cdot 10^4$ (K$^{-1/2}$) at $T \cong 300$ K | 1.6 | 2.3 | 1.6 | 3.0 | 4.0 |

(*) - twist, (**) – Dewar

During weighing, some various physical factors were considered: convection, buoyancy, action of heat and ultrasound on the balance, influence of magnetic and electric fields, and others. According to quantitative estimations, the accuracy of the temperature factors $\alpha$ was about 20-25 %.

The authors also constructed a thermo-physical model of reduction of the apparent weight of non-uniformly heated samples which model agreed well with the experiment. It is typical that the negative temperature dependence of body weight is always observed, with the greatest values of factor $\alpha$ being obtained for the samples made of light and elastic materials. Let's note that for the first time the negative temperature dependence of weight of bodies was actually observed in experiments of Show and Davy described in 1923 (Shaw and Davy, 1923); however, the authors then did not dare to insist on their results (Dmitriev, 2006).

## MEASUREMENTS OF ANISOTROPY OF RUTILE CRYSTAL WEIGHT

The elementary analysis shows that the acceleration of microparticles of a test body in their thermal movement is directly proportional to $\sqrt{c/\mu}$, where c is the factor of elasticity and $\mu$ is the mass of the particles. If V is the speed of elastic longitudinal waves in a considered body and $\rho$ the density of the body's material, then the following ratio is true

$$a \propto \sqrt{c}/\mu \propto V/\sqrt{\rho} \quad . \tag{7}$$

The consequence of Equ. (7) should be the dependence of weight of an anisotropic crystal on its orientation (Dmitriev and Chesnokov, 2004).

If the speeds $V_1$, $V_2$ of longitudinal waves in a crystal for orthogonal directions noticeably differ, then at the constant crystal temperature the relative difference $\Delta P/P$ of its weight measured in two positions "1" and "2" is equal to

$$\frac{\Delta P}{P} \propto -\frac{V_1 - V_2}{\sqrt{\rho}} \quad . \tag{8}$$

The experimental results of measurement of weight differences of a rutile crystal measured at its mutual - perpendicular positions are given in Fig. 6.

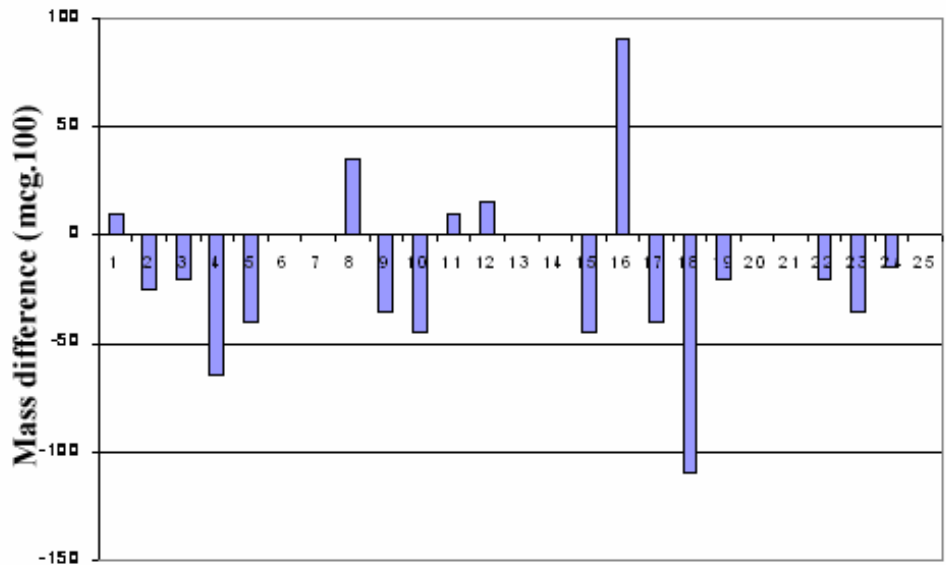

**FIGURE 6.** Weight differences of rutile crystal measured at its mutual - perpendicular positions; one division of a ordinate-scale corresponds to 0.5 mcg; results of four series of measurements are shown.

The average weight of a sample is 2.876 g, speeds of longitudinal sound waves are $V_1=10.94 \cdot 10^3$ m/s and $V_2=8.014 \cdot 10^3$ m/s. The relative difference of weights is equal to $-7 \cdot 10^{-8}$ with root-mean-square deviation of 0.10 mcg. Big fluctuations of measured differences of weights are caused by instability of temperature regimes of weighing. Nevertheless, there prevails the regular character of inequality of weight observed in case of anisotropic crystal the sign of which corresponds to the ratio in Equ. (8).

## CONCLUSION

So, the four described experiments show that the accelerated movement of a body caused by action of elastic (electromagnetic in nature) forces influences the force of its gravitational interaction with others, conditionally motionless, bodies. Indirectly, such an influence causes a negative temperature dependence of body weight that has a big practical value for precision gravimetry, for fundamental problems of physics of gravitation, and also in interpretation of some phenomena of astrophysics (Dmitriev, 2005; Dmitriev, 2006a, 2006b).

In the immediate prospects, it seems necessary to conduct the following experimental research

- precision measurement of physical temperature dependence of weight of various materials in a wide range of temperatures,

- dynamic measurements of weight of bodies in a condition of elastic effects: acoustic (including, ultra-and hypersonic) and impact, and also while their oscillatory and rotary movements,

- measurement of a mutual attraction forces of accelerated moving weights at the lowest temperatures.

Experimental and theoretical research of problems of interaction of acceleration and gravitation, and also the problems of temperature dependence of forces of gravitation connected with them, has rather big value both for the development of physics of gravitation and for perspective technologies of the future.

# REFERENCES

Dmitriev A. L., "On the Influence of External Elastic (Electromagnetic) Forces on the Gravity," *Russian Physics Journal,* **44**(12), 1323-1327, (2001).

Dmitriev A. L., Snegov V. S., "Weighing of the Mechanical Gyros with Horizontal and Vertical Orientations of Axis of Rotation," *Izmeritelnaja Tekhnika*, **8**, 33 – 35 (2001), (in Russian).

Dmitriev A. L., "Inequality of the Coefficients of Restitution for Vertical and Horizontal Quasielastic Impacts of a Ball Against a Massive Plate," *International Applied Mechanics*, **38**(6), 747 – 749, (2002).

Dmitriev A. L., Nikushchenko E. M., Snegov V. S., "Influence of the Temperature of a Body on its Weight," *Measurement Techniques*, **46**(2), 115 – 120, (2003).

Dmitriev A. L., Chesnokov N. N., "Influence of Orientation of Anisotropic Crystal on its Weight," *Izmeritelnaja Tekhnika*, **9**, 36 – 37, 2004, (in Russian).

Dmitriev A. L., "*Upravljaemaja Gravitacija,*" Novy Centr, Moscow, 2005, 70 p., (in Russian).

Dmitriev A. L., "On Possible Causes of Divergence in Experimental Values of Gravitation Constant," http://arxiv.org/physics/0610282, accessed November 1, 2006.

Dmitriev A. L., "Temperature Dependence of Gravitational Force: Experiments, Astrophysics, Perspectives," http://arxiv.org/physics/0611173, accessed November 20, 2006.

Shaw P. E., Davy N., "The Effect of Temperature on Gravitative Attraction," *Phys. Rev*., **21**(6), 680 – 691, (1923).

Tajmar M., Plesescu F., Seifert B., Schnitzer R., Vasiljevich I., "Search for Frame-Dragging in the Vicinity of Spinning Superconductors," http://arxiv/org/gr-gc/07073806, accessed July 25, 2007.